\documentclass[]{aa}
\usepackage{graphicx}
\usepackage{times}
\usepackage{txfonts}
\usepackage{epsfig,latexsym,longtable,lscape}
\usepackage{natbib}
\bibpunct{(}{)}{;}{a}{}{,}

\newcommand{\SII}{[S\,{\sc ii}]}
\newcommand{\OIII}{[O\,{\sc iii}]}
\newcommand{\Halpha}{H${\alpha}$}
\newcommand{\D}{$^\circ$}
\def\p0{\phantom{0}}

\def\SNR{\mbox{{SNR~J0530--7007}}}

\begin{document}
\title{Multi-Frequency Study of Supernova Remnants in the Large Magellanic Cloud.}
\subtitle{The case of LMC \SNR}
\author{A.Y. De Horta\inst{1}
\and M.D.~Filipovi\'c\inst{1}
\and L.M.~Bozzetto\inst{1}
\and P.~Maggi\inst{2}
\and F.~Haberl\inst{2}
\and E.J.~Crawford\inst{1}
\and M.~Sasaki\inst{3}
\and D.~Uro{\v s}evi{\'c}\inst{4,5}
\and W.~Pietsch\inst{2}
\and R.~Gruendl\inst{6}
\and J. Dickel\inst{7} 
\and N.F.H.~Tothill\inst{1}
\and Y.-H.~Chu\inst{6} 
\and J.L.~Payne\inst{1}
\and J.D.~Collier\inst{1}
 }
\institute{
School of Computing and Mathematics, University of Western Sydney Locked Bag 1797, Penrith South DC, NSW 1797, Australia
\and 
Max-Planck-Institut f\"{u}r extraterrestrische Physik, Giessenbachstra\ss e, D-85748 Garching, Germany
\and
Institut f\"ur Astronomie und Astrophysik T\"ubingen, Sand 1, D-72076 T\"ubingen, Germany
\and
Department of Astronomy, Faculty of Mathematics, University of Belgrade, Studentski trg 16, 11000 Belgrade, Serbia
\and
Isaac Newton Institute of Chile, Yugoslavia Branch 
\and
Department of Astronomy, University of Illinois, 1002 West Green Street, Urbana, IL 61801, USA
\and
Physics and Astronomy Department, University of New Mexico, MSC 07-4220, Albuquerque, NM 87131, USA 
}


\abstract
 {The Supernova Remnants (SNRs) known in the Large Magellanic Cloud (LMC) show a variety of morphological structures in the different wavelength bands. This variety is the product of the conditions in the surrounding medium with which the remnant interacts and the inherent circumstances of the supernova event itself.}
 { This paper performs a multi-frequency study of the LMC \SNR\ by combining Australia Telescope Compact Array (ATCA), Molonglo Observatory Synthesis Telescope (MOST), R\"ontgensatellit (\emph{ROSAT}) and Magellanic Clouds Emission Line Survey (MCELS) observations.}
 {We analysed radio-continuum, X-ray and optical data and present a multi-wavelength morphological study of LMC \SNR.}
 {We find that this object has a shell-type morphology with a size of 215\arcsec$\times$180\arcsec\ (52~pc $\times\ $44~pc); a radio spectral index ($\alpha=-0.85\pm0.13$);  with \SII/\Halpha$>$0.4 in the optical; and the presence of non-thermal radio and X-ray emission.}
 {We confirmed this object as a bona-fide shell-type SNR which is probably a result of a Type~Ia supernova.}


\maketitle


\section{Introduction}

Lying towards the South Ecliptic Pole, the Large Magellanic Cloud (LMC), is in one of the coldest parts of the radio sky, uncontaminated by Galactic foreground emission \citep{1991A&A...252..475H}. The LMC's position and its known distance of 50\,kpc \citep{2008MNRAS.390.1762D} makes the LMC arguably the best galaxy in which to study supernova remnants (SNRs) in our Local Group of galaxies.

In the radio-continuum, SNR emission is predominantly non-thermal, giving rise to a typical radio spectral index of $\alpha\sim-0.5$ ($S\propto\nu^\alpha$). However, the environment in which the SNR evolves i.e.\ the interstellar medium (ISM) with its ambient magnetic field, will not only affect the radio spectral index observed but also the SNR's morphology, structure and behaviour \citep{1998A&AS..130..421F}.

In a H${\alpha}$ survey of the LMC, \citet{1976MmRAS..81...89D} reported ``diffuse filaments" with a size of 10\arcmin$\times$9\arcmin\ at \mbox{RA~(J2000)=$5^\mathrm{h}30^\mathrm{m}$30\fs35} and \mbox{Dec~(J2000)=--70\degr07\arcmin51\farcs5} and named it DEM\,L218. A radio source designated 0531-701 was identified by \citet{1984PASAu...5..537T} at \mbox{RA~(J2000)=$5^\mathrm{h}30^\mathrm{m}$38\fs19} and \mbox{Dec~(J2000)=--70\degr07\arcmin33\arcsec}  and classified as an SNR candidate in a survey with the Molonglo Observatory Synthesis Telescope (MOST). \citet{1998A&AS..130..421F} detected this source in the Parkes radio surveys of the Magellanic Clouds (MCs; $\lambda$=6~cm~and~3~cm) but due to a rather flat spectrum of \mbox{$\alpha$=--0.17$\pm$0.24} and the survey's low resolution (Parkes Beam Sizes: 4\farcm9 at $\lambda$=6~cm; 2\farcm7 at $\lambda$=3~cm), they were unable to classify it as an SNR at that time. \citet[][hereafter HP99]{1999A&AS..139..277H}\defcitealias{1999A&AS..139..277H}{HP99} detected a nearby R\"ontgensatellit (\emph{ROSAT}) X-ray source ([HP99]\,1081) at a position of RA~(J2000)=05$^h$30$^m$51.8$^s$ and Dec~(J2000)=$-$70\degr06\arcmin44\arcsec; however, this object is close to the LMC bar where confusion is significant. The object was re-discovered using the Magellanic Clouds Emission Line Survey (MCELS) \citep{2004AAS...20510108S}. This study inferred that the object is likely a large, old radiative shell-type SNR with enhanced \SII\ (\SII/\Halpha$>$0.4). \citet{2006ApJS..165..480B} also observed the object using the Far Ultraviolet Spectroscopic Explorer (FUSE) satellite, at a position of \mbox{RA(J2000)=05$^h$30$^m$37$^s$}, \mbox{Dec(J2000)=--70\degr08\arcmin40\arcsec} with a beamsize of 145\arcsec. Weak, yet moderately broad lines of O\,{\sc vi} were detected, in addition to possible, but uncertain C\,{\sc iii} lines. \citet{2010AJ....140..584D} reported that there are no molecular clouds detected towards this object and that there are no young stellar objects in its vicinity.

In this paper, we report new Australia Telescope Compact Array (ATCA) radio-continuum observations at at $\lambda$=3~cm and 6~cm. These new radio-continuum observations in conjunction with previous radio-continuum ($\lambda$=20~cm and~13~cm (ATCA),  $\lambda$=36~cm (MOST)), X-ray (\emph{ROSAT}) and optical (MCELS) observations are used to confirm that the object in the LMC centered at \mbox{RA~(J2000)=5$^h$30$^m$40.4$^s$} and \mbox{Dec~(J2000)=--70\degr07\arcmin27.4\arcsec} is a bona-fide SNR that, hereafter, we will call \SNR. The observations, data reduction and imaging techniques are described in Sect.~\ref{section:observations}. Astrophysical interpretation of the newly obtained moderate-resolution total intensity images, in combination with existing \emph{ROSAT} and MCELS images are discussed in Sect.~\ref{section:rad}.


\section{Observations and data reduction}
 \label{section:observations}

 \subsection{Radio-continuum}
   \label{datareduction_radio}

Radio-continuum observations at the five frequencies shown in Table~\ref{tbl-1} have been used to study and measure the flux densities of \SNR. For the 36~cm (MOST) flux density measurement given in Table~\ref{tbl-1}, we used the unpublished image as described by \citet{1984AuJPh..37..321M}. The 20~cm (ATCA) image used is from \citet{2007MNRAS.382..543H}. 

\begin{table}
 \caption{Integrated flux densities of \SNR.}
 \label{tbl-1}
 \centering
 \begin{tabular}{ccccc}
  \hline\hline
 $\nu$ & $\lambda$ & Beam Size        & R.M.S & S$_\mathrm{Total}$ \\
 (MHz) & (cm)      & (\arcsec)        & (mJy) & (mJy)\\
  \hline
 \p0843& 36        & 43.0$\times$43.0 & 0.5\p0& 107\tablefootmark{a}\\
 \p0843& 36        & 43.0$\times$43.0 & 0.5\p0& \p080\\
 1400  & 20        & 40.0$\times$40.0 & 0.5\p0& \p062\\
 2400  & 13        & 54.1$\times$48.9 & 0.4\p0& \p052\\
 5500  & \p06      & 33.8$\times$33.8 & 0.05  & \p023\\
 9000  & \p03      & 22.5$\times$22.5 & 0.05  & \p010\\
  \hline
 \end{tabular}
 \tablefoot{
 \tablefoottext{a} Integrated flux density from \citet{1984PASAu...5..537T}.
}
\end{table}

\begin{figure}[h]
 \hspace*{-10 mm}\includegraphics[angle=-90,width=.63\textwidth]{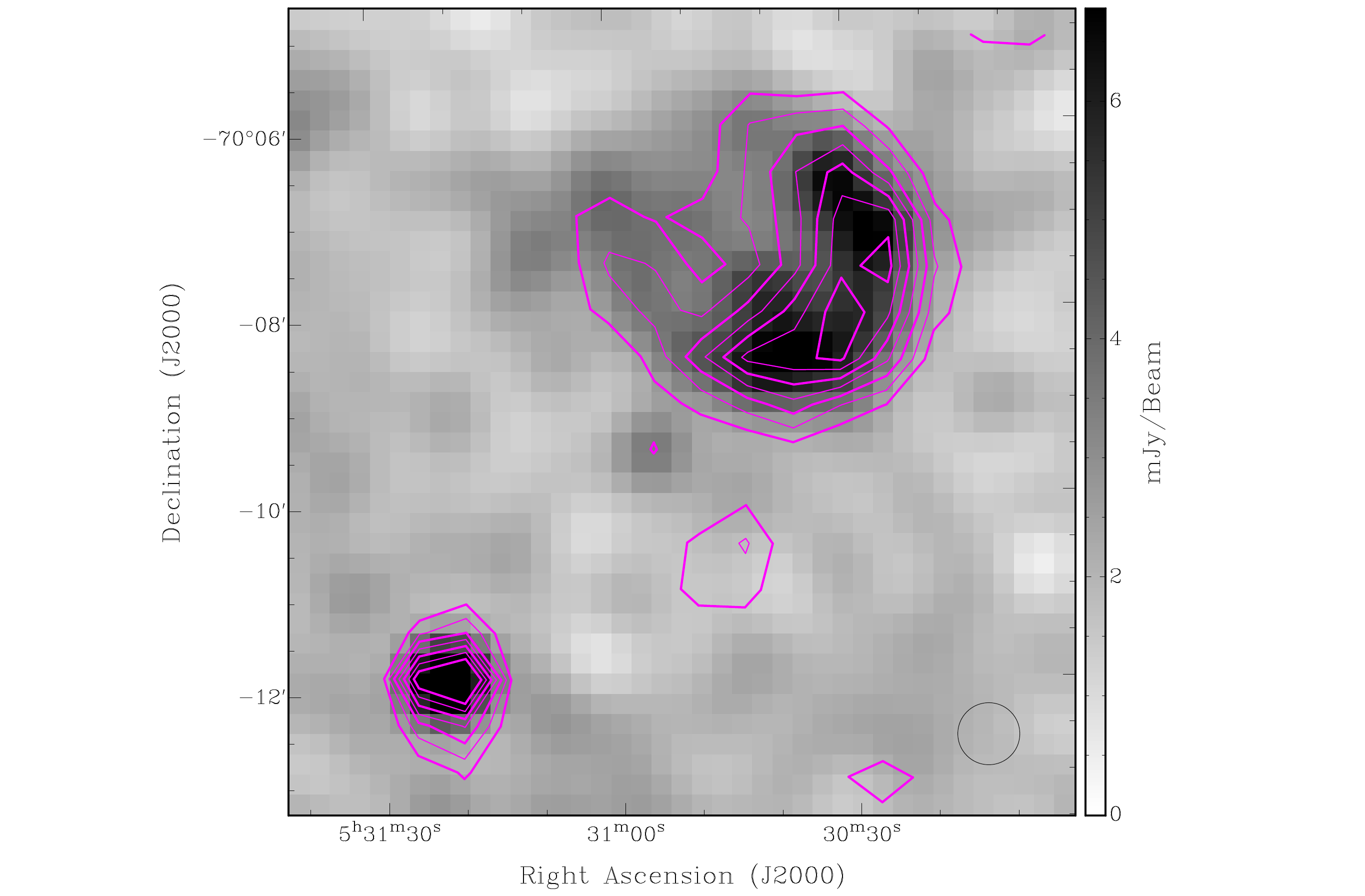}
 \caption{Combined ATCA observations of \SNR\ at 20~cm (1400~MHz) overlaid with MOST 36~cm (843~MHz) contours. Contours are from 2 to 8~mJy/beam in steps of 1~mJy/beam. The black circle in the lower right corner represents the synthesised beamwidth (at 20~cm) of 40\arcsec.}
 \label{fig1}
\end{figure}

Australia Telescope Compact Array (ATCA) observations from 1992 April 29 and 30  (project C195) of \SNR\ were also used. These observations were made at wavelengths $\lambda$=13~cm and 20~cm ($\nu$=2400\,MHz and 1400\,MHz) using the 375 array configuration. 

Recently, on 2011 November 15, we observed \SNR\  (project C634), with the ATCA in the EW352 configuration, at $\lambda$=3/6~cm (9000 and 5500\,MHz) and a bandwidth of 2\,GHz (ATCA project C634). In addition to this, \SNR\ was observed using the ATCA on 1995 November 12 with the 6A array  (ATCA Project C461; $\lambda$=3/6~cm ($\nu$=8640\,MHz and 4800\,MHz)). Both observations were carried out in snap-shot mode, totalling about 1 hr (each) of integration over a 12 hour period. These two sets of observations were combined with mosaic observations from project C918 \citep{2005AJ....129..790D}. 

For all these radio-continuum observations, baselines formed with ATCA antenna 6 were excluded, as the other five antennas were arranged in a compact configuration. Source \mbox{PKS~B1934--638} was used for primary (flux) calibration and source \mbox{PKS~B0530--727} was used for secondary (phase) calibration. More information on the observing procedure and other sources observed under the project C634 can be found in \citet{2007MNRAS.378.1237B,2008SerAJ.177...61C,2008SerAJ.176...59C,2009SerAJ.179...55C,2010A&A...518A..35C,2010SerAJ.181...43B,2011arXiv1109.3945B,2012MNRAS.tmp.2167B}. Parkes radio-continuum data from \citet{1995A&AS..111..311F,1996A&AS..120...77F} were initially combined with the ATCA data wherever the shortest ATCA baselines were less than the Parkes diameter of 64\,m, and the observational frequencies corresponded, in order to provide zero-spacing. However, as the R.M.S noise of the Parkes observations are significantly higher, especially at 3/6~cm, it was decided not to use these observations in our final analysis.

\begin{figure}
 \hspace*{-10 mm}\includegraphics[angle=-90,trim=20 75 0 0,width=.67\textwidth]{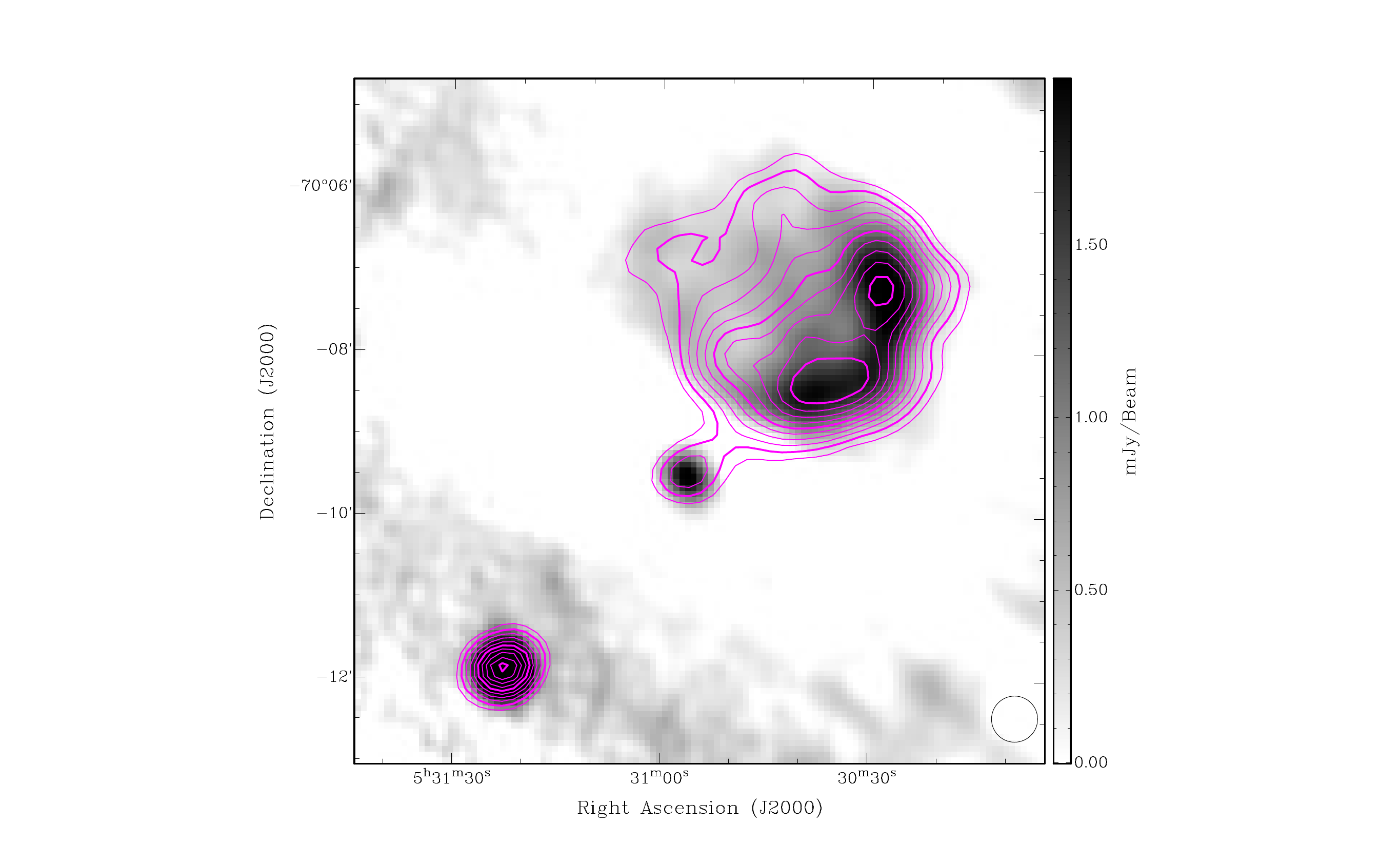}
 \caption{Combined ATCA observations of \SNR\ at 6~cm (5500~MHz) overlaid with 13~cm (2400~MHz) contours. Contours are from 1.5 to 6~mJy/beam in steps of 0.5~mJy/beam. The black circle in the lower right corner represents the synthesised beamwidth (at 6~cm) of 33.8\arcsec.}
 \label{fig2}
\end{figure}

The total-intensity images in Fig.~\ref{fig1} and Fig.~\ref{fig2} were formed using the standard \textsc{miriad}  \citep{1995ASPC...77..433S} tasks employing multi-frequency synthesis using a natural weighting scheme with a correction for the primary beam response applied. A similar procedure was used for both \textit{U} and \textit{Q} Stokes parameter maps. However, due to the low dynamic-range (signal to noise ratio between the source flux and $3\sigma$ noise level) self-calibration could not be applied. 

\subsection{Optical}
 \label{datareduction_optical}

\begin{figure}
\includegraphics[angle=-90,width=0.45\textwidth]{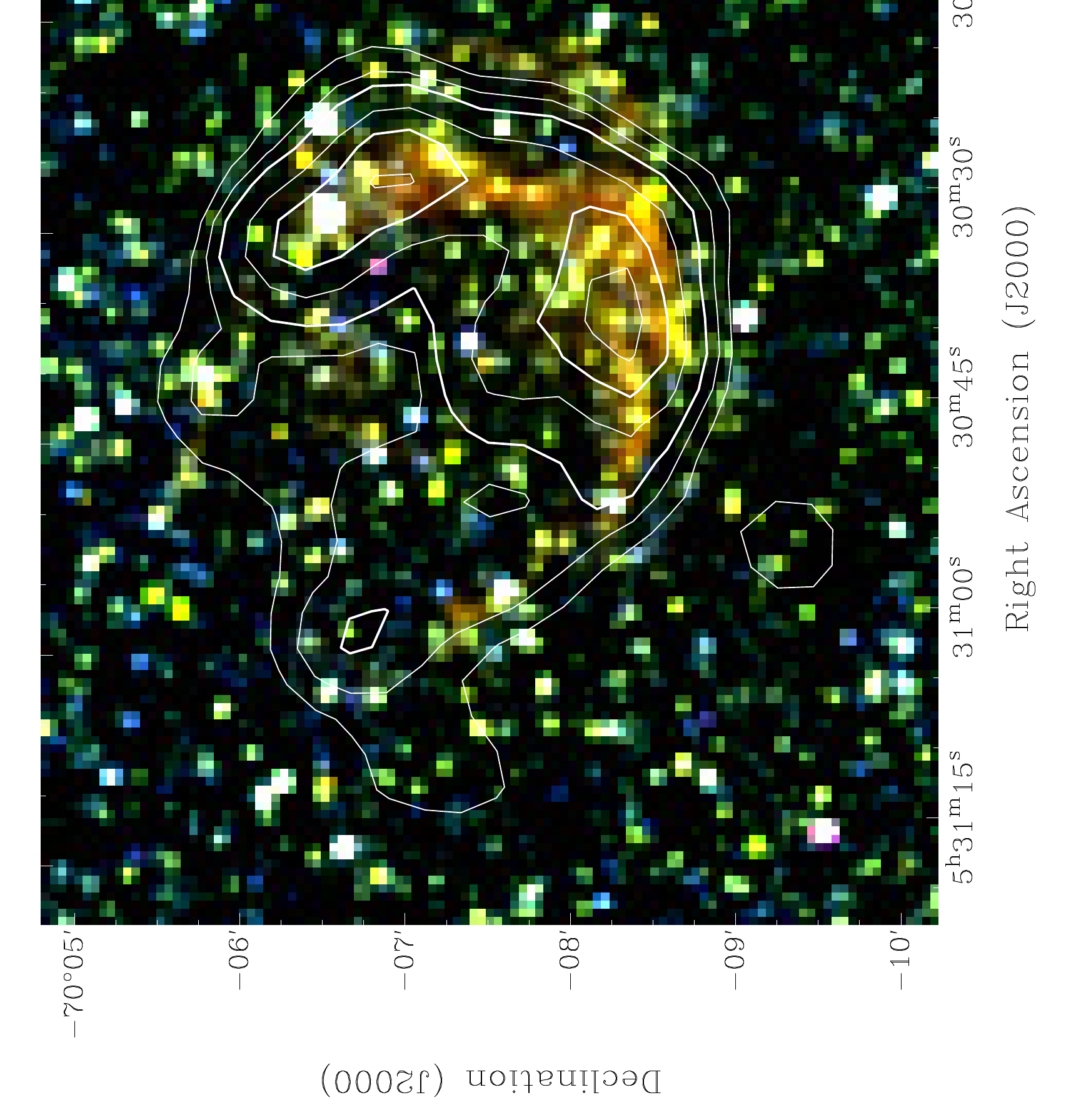}
\caption{MCELS composite optical image \textrm{(RGB = H$\alpha$,[S\textsc{ii}],[O\textsc{iii}])} of \SNR\ overlaid with 20~cm radio contours. Contours are 3, 3.5, 4, 5, 6 and 7~mJy/beam. }
 \label{fig3}
\end{figure}

The MCELS observations \citep{2006NOAONL.85..6S} were carried out with the 0.6~m University of Michigan/Cerro Tololo Inter-American Observatory (CTIO) Curtis Schmidt telescope, equipped with a SITe $2048 \times 2048$\ CCD, giving a field of 1.35\degr\ $\times$ 1.35\degr\ at a pixel scale of 2.4\arcsec\ $\times$ 2.4\arcsec. It mapped both the LMC and SMC in narrow bands covering \Halpha, \OIII\ ($\lambda$=5007\,\AA), and \SII\ ($\lambda$=6716,\,6731\,\AA). Also observed were matched red and green continuum bands, used primarily to subtract the stars from the images to reveal the full extent of the faint diffuse emission. All the data has been flux-calibrated and assembled into mosaic images; a small section of the mosaic is shown in Fig.~\ref{fig3}. 

The high-resolution H$\alpha$ image in Fig.~\ref{fig3} was obtained with the MOSAIC~II camera on the Blanco 4-m telescope at the CTIO. It confirms a distinctive optical nebulosity associated with the SNR candidate. Here, for the first time, we present optical images of this object in combination with our new ATCA radio-continuum and \emph{ROSAT} X-ray data.

\subsection{X-rays}
 \label{datareduction_Xray}
 
\begin{figure}
 \begin{center}
\includegraphics[scale=0.47]{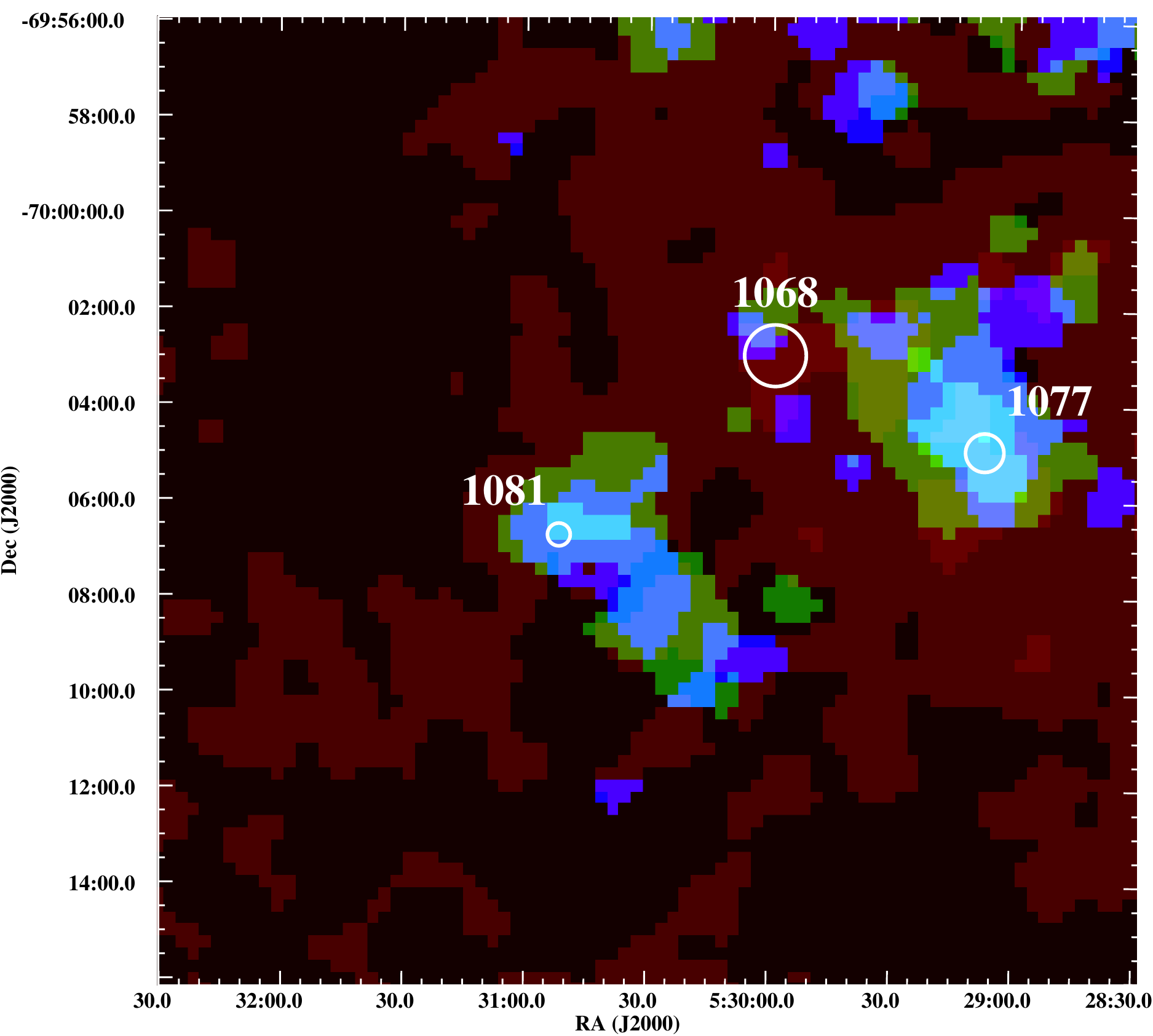}
 \caption{The \emph{ROSAT} PSPC RGB colour image of the area around \SNR. The energy bands are: red (0.1--0.4 keV), green (0.5--0.9~keV) and blue (0.9--2.0~keV). The image has a pixel size of 15\arcsec\ and is smoothed with a $\sigma$ of 1.5~pixel. The annotations denote sources from \citetalias{1999A&AS..139..277H}.}
 \label{fig4}
 \end{center}
\end{figure}

\begin{figure}
 \begin{center}
\includegraphics[angle=-90,scale=0.49]{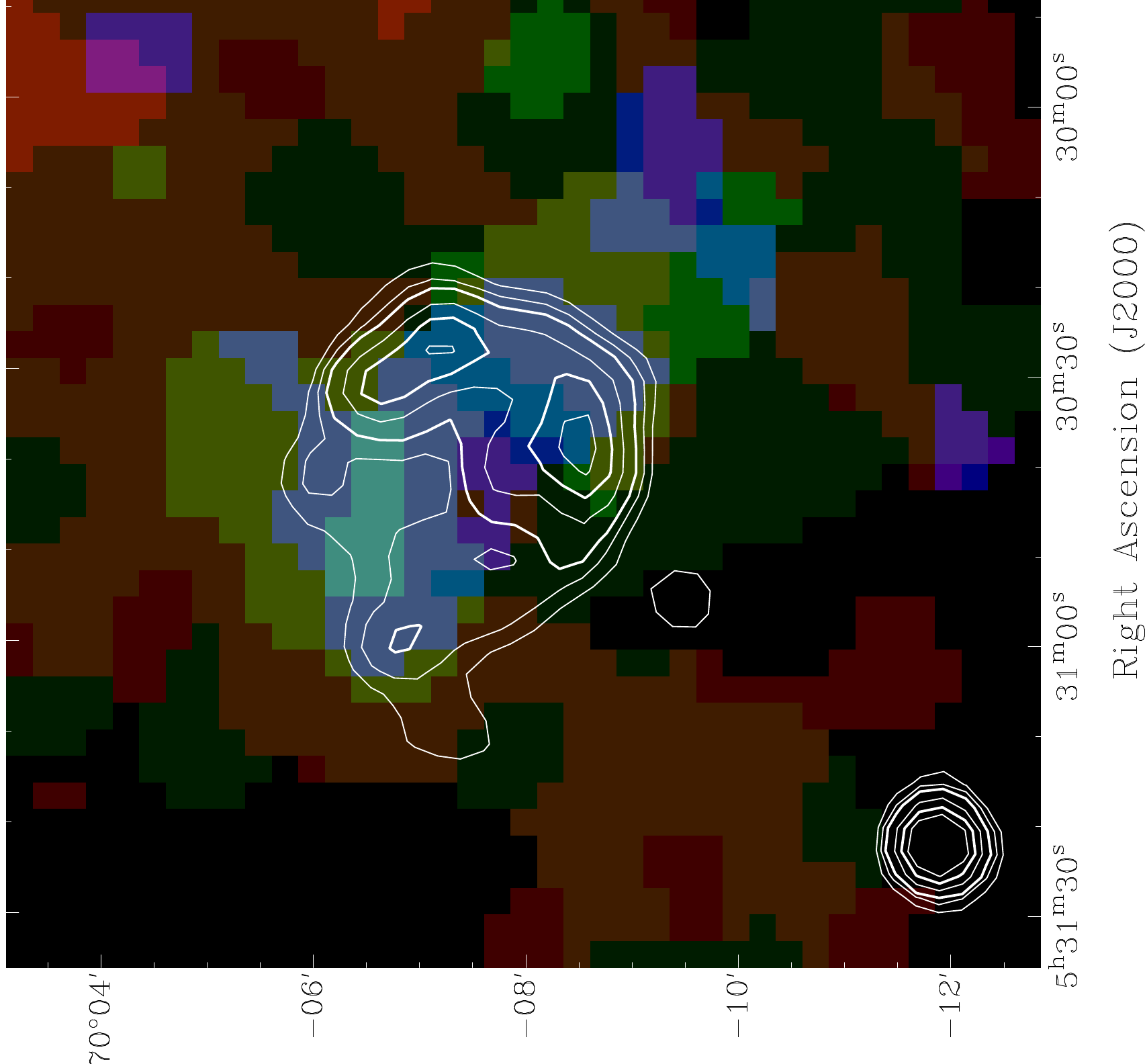}
 \caption{The \emph{ROSAT} PSPC RGB colour image of \SNR\ overlaid with the contours of 20~cm intensity. Contours are 3, 3.5, 4, 5, 6 and 7~mJy/beam. }
 \label{fig5}
 \end{center}
\end{figure}

\SNR\ lies in the field of view of several pointed observations of \emph{ROSAT}'s Position Sensitive Proportional Counter (PSPC). The \citetalias{1999A&AS..139..277H} catalogue includes a very weak detection ([HP99]\,1081, Fig.~\ref{fig4}) within the extent of the radio emission from \SNR. However, the X-ray properties of this source are unclear: The hardness ratios are poorly defined (HR1=1.00$\pm$1.23) or undefined (HR2) and therefore give no meaningful information on the spectrum. The \citetalias{1999A&AS..139..277H} catalogue was derived from  individual PSPC observations, without combining the exposures of overlapping fields. To investigate [HP99]\,1081 and its possible association with \SNR\ in more detail, we selected 9 observations of the LMC which covered the SNR within 24\arcmin\ of the optical axis (to avoid the degraded point spread function at larger off-axis angles). In Table~\ref{ROSATobs_full} we give the \emph{ROSAT} sequence number, target name, exposure time and central coordinates of the selected pointings, as well as the off-axis angle of \SNR\ in each of them. Images were produced at different energy bands (broad: 0.1--2.4~keV, soft: 0.1--0.4~keV, hard: 0.5--2.0~keV, hard1: 0.5--0.9~keV and hard2: 0.9--2.0~keV) from the merged data. A colour image of the area around the SNR with net exposure (vignetting corrected) of $\sim$48~ks is shown in Figs.~\ref{fig4} and \ref{fig5} with red, green and blue representing the X-ray intensities in the soft, hard1 and hard2 bands. The resolution of the \emph{ROSAT} PSPC varies with energy but the point spread function is always less than 1\arcmin.

The \citetalias{1999A&AS..139..277H} catalogue contains two other sources detected in the neighbourhood of source 1081 (see Fig.~4). [HP99]\,1068 is a weak source with an existence likelihood of 11.8, just above the threshold used for the catalogue. No useful information about extent or hardness ratios can be derived. Source [HP99]\,1077 is a clear detection (existence likelihood 22.8) with indication for an extent of 31\arcsec (likelihood for the extent of 13.3). Therefore, [HP99]\,1077 looks like an extended source itself, but no significant radio nor optical emission in the MCELS images is seen at its position. Future X-ray observations are required to investigate the whole region around the three ROSAT sources in more detail. 

\begin{table*}[t]
\smallskip
\caption{\emph{ROSAT} observations summary of \SNR\ (sorted by RA).}
\begin{center}
\begin{tabular}{l l c c c c}   
  \hline
    \noalign{\smallskip}
\emph{ROSAT} & Target name  & Obs. time & RA                 & DEC                           & Off-axis angle\tablefootmark{a} \\
 sequence &                 &(sec) & \multicolumn{2}{c}{ (J2000)}                            & (arcmin)\\
\noalign{\smallskip}
\hline
\noalign{\smallskip}
180287p   & NOVA LMC 1995     & 2531 & 05$^h$26$^m$50.04$^s$ & --70\degr01\arcmin11.5\arcsec & 21.3 \\      
400298p   & RX\,J0527.8--6954 & 1058 & 05$^h$27$^m$48.00$^s$ & --69\degr54\arcmin00.0\arcsec & 20.2 \\      
400298p-1 & RX\,J0527.8--6954 & 7502 & 05$^h$27$^m$48.00$^s$ & --69\degr54\arcmin00.0\arcsec & 20.2 \\    
400298p-2 & RX\,J0527.8--6954 & 7802 & 05$^h$27$^m$48.00$^s$ & --69\degr54\arcmin00.0\arcsec & 20.2 \\
400148p   & RX\,J0527.8--6954 & 6064 & 05$^h$27$^m$48.00$^s$ & --69\degr54\arcmin00.0\arcsec & 20.2 \\      
180255p   & RX\,J0527.8--6954 & 9763 & 05$^h$27$^m$50.04$^s$ & --69\degr54\arcmin00.0\arcsec & 20.0 \\
300172p   & NOVA LMC 1988 a   & 6272 & 05$^h$32$^m$28.08$^s$ & --70\degr21\arcmin36.0\arcsec & 17.0 \\
300172p-1 & NOVA LMC 1988 a   & 2993 & 05$^h$32$^m$28.08$^s$ & --70\degr21\arcmin36.0\arcsec & 17.0 \\
300172p-2 & NOVA LMC 1988 a   & 3880 & 05$^h$32$^m$28.08$^s$ & --70\degr21\arcmin36.0\arcsec & 17.0 \\   
\noalign{\smallskip}
\hline
  \end{tabular}
  \tablefoot{
  \tablefoottext{a} Mean angular distance of [HP99] 1081 to the optical axis of the telescope.
  }
 \label{ROSATobs_full}
 \end{center}
\end{table*}

\section{Results and Discussion}
 \label{section:rad}

The remnant has a typical horse-shoe morphology (Figs.~\ref{fig1}, \ref{fig2} and \ref{fig3}), centered at \mbox{RA~(J2000)=5$^h$30$^m$40.4$^s$} and \mbox{Dec~(J2000)=--70\degr07\arcmin27.4\arcsec}. The measured position differs from that of the FUSE observations \citep{2006ApJS..165..480B}, since the FUSE observation was aimed only at the Southern side of the shell and not at the centre of the SNR. 

The size of \SNR\  at $\lambda=20$~cm is 215\arcsec$\pm$4\arcsec\ $\times$ 180\arcsec$\pm$4\arcsec\ \mbox{(52$\pm$1 pc $\times$ 44$\pm$1 pc)}. The size  was measured by taking line profiles along the major (NE--SW) and minor (SE--NW) axis (PA=45\D) of the remnant using the \textsc{karma}\footnote{http://www.atnf.csiro.au/computing/software/karma/} \citep{2006Karma} tool {\sc kpvslice} and detemining the distance between the point when the line profile first rises above 3$\sigma$ (1.50~mJy) and the point when it finally falls below 3$\sigma$. The thickness of the shell is estimated to be $<$30\arcsec\ (7~pc) at 6~cm, about 30\% of the SNR's radius (ie. a filling factor of 0.64). 

The merged \emph{ROSAT} PSPC images reveal an elongated structure of X-ray emission at the location of the SNR with a brighter spot right at the position of [HP99]\,1081. The presence of extended X-ray emission is coincident with the radio emission of the SNR (Fig.~\ref{fig5}). In X-rays, the brightest part is in the north-east while the MCELS optical emission closely follows radio-continuum appearance (Fig.~\ref{fig3}). 

\begin{figure}[h]
  \includegraphics[angle=-90, width=\columnwidth]{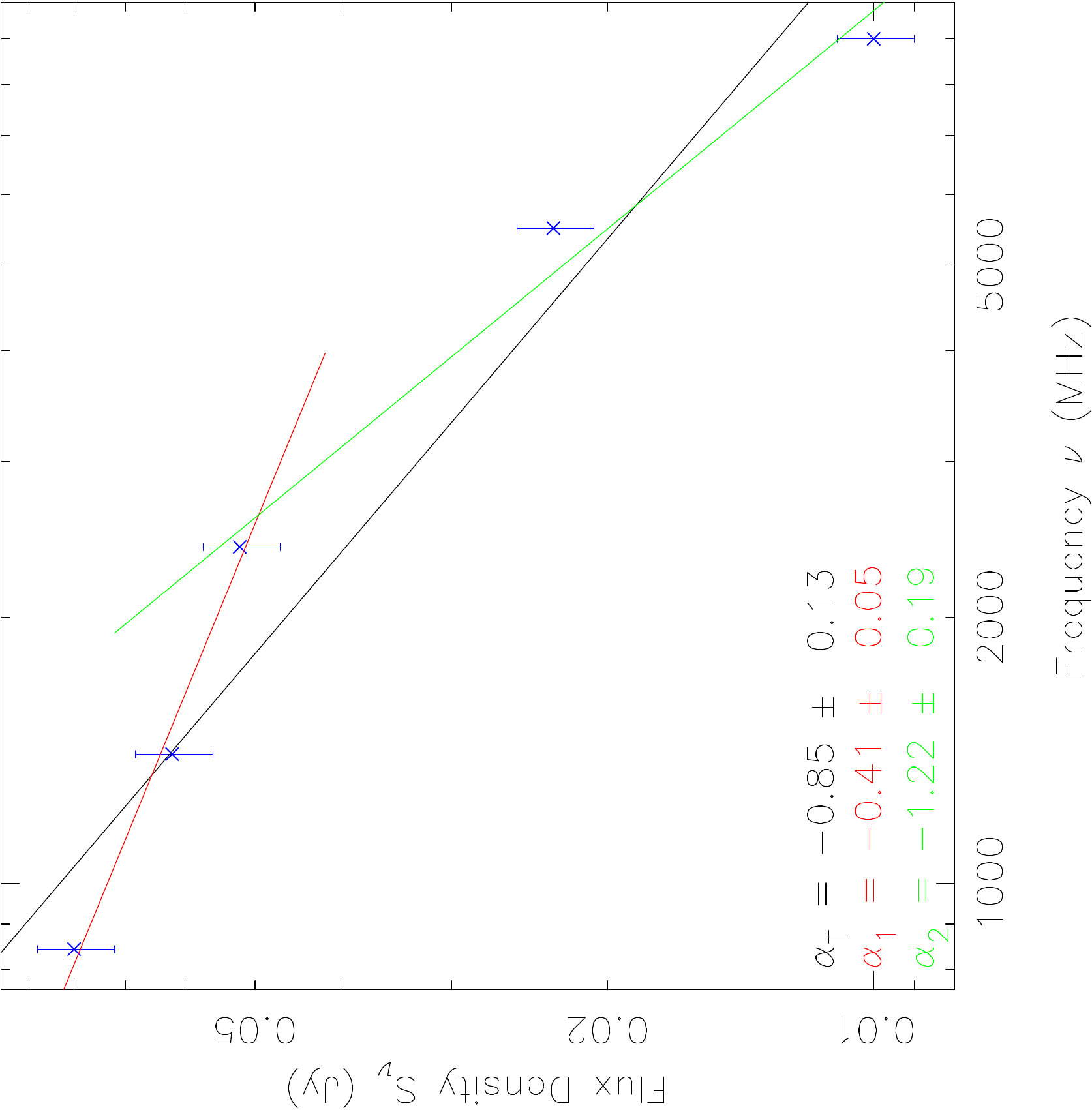}
  \caption{Radio-continuum spectrum of \SNR. The black line ($\alpha_{\mathrm{T}}$) is the overall radio-continuum spectra, the red line ($\alpha_{1}$) between 36~cm~and 13~cm, and the green line ($\alpha_{2}$) between 13~cm~and~3~cm. Note log-log scale.}
  \label{fig6}
\end{figure}

\begin{figure}[h]
 \includegraphics[width=0.5\textwidth]{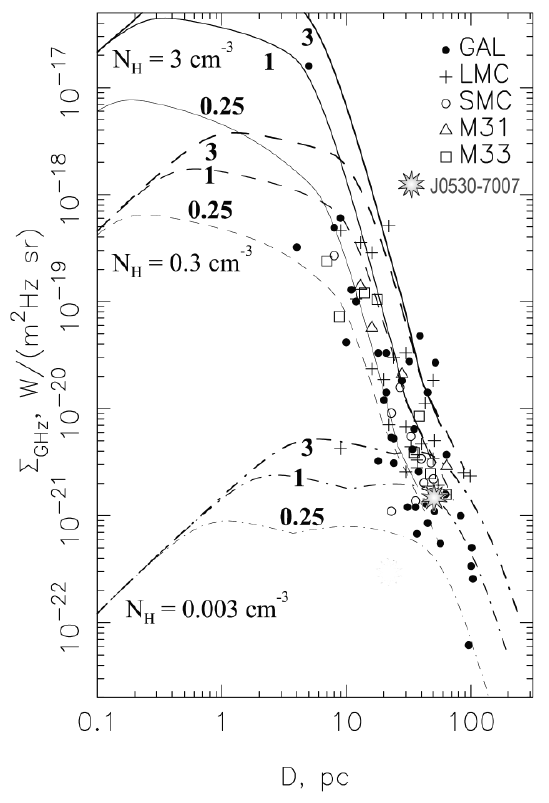}
 \caption{1 GHz Surface brightness-to-diameter diagram from \citet{2004A&A...427..525B}, with \SNR\ added. The evolutionary tracks are for ISM densities of N$_\mathrm{H}$= 3, 0.3 and 0.003~cm$^{-3}$ and explosion energies of E$_{\mathrm{SN}}$ = 0.25, 1 and 3$\times10^{51}$~erg.}
 \label{fig7}
\end{figure}

We note that the significant difference in the flux density measurement at 36~cm (107~mJy in \cite{1984PASAu...5..537T} vs 80~mJy in this work) may introduce a very large uncertainty in the spectral index measurement ($S\propto\nu^{\alpha}$). A possible explanation for this discrepancy is that we applied different fitting model then \cite{1984PASAu...5..537T}. Using all values of integrated flux density estimates (except for 36~cm value from \cite{1984PASAu...5..537T}; Table~1), a spectral index ($S\propto\nu^{\alpha}$) distribution is plotted in Fig.~\ref{fig6}. The overall radio-continuum spectra (Fig.~\ref{fig6}; black line) from \SNR\ was estimated to be $\alpha_{\mathrm{T}}=-0.85\pm0.13$, while the typical SNR spectral index is $\alpha=-0.5\pm0.2$ \citep{1998A&AS..130..421F}. This somewhat steeper spectral index would indicate a younger age despite its (large) size of $52\times$44~pc, suggesting it as an older (more evolved) SNR. We also note that this may indicate that a simple model does not accurately describe the data, and that a higher order model is needed. This is not unusual, given that several other Magellanic Clouds SNR's exhibit this ``curved'' spectra \citep{2008SerAJ.177...61C,2010SerAJ.181...43B,2011arXiv1109.3945B}. Noting the breakdown of the power law fit at shorter wavelengths, we decomposed the spectral index estimate into two components, one ($\alpha_{1}$) between 36 and 13~cm, and the other ($\alpha_{2}$) between 13 and 3~cm. The first component (Fig.~\ref{fig6}; red line), $\alpha_{1}=-0.41\pm0.05$ is a reasonable fit and typical for an evolved SNR, whereas the second (Fig.~\ref{fig6}; green line), \mbox{$\alpha_{2}=-1.22\pm0.19$,} is a poor fit, and indicates that non-thermal emission can be described by different populations of electrons with different energy indices. Although the low flux at 3~cm (and to a lesser extent at 6~cm) could cause the large deviations, an underestimate of up to $\sim$50\% would still lead to a ``curved'' spectrum.

\begin{figure*}[ht]
 \begin{center}
\includegraphics[scale=1.0]{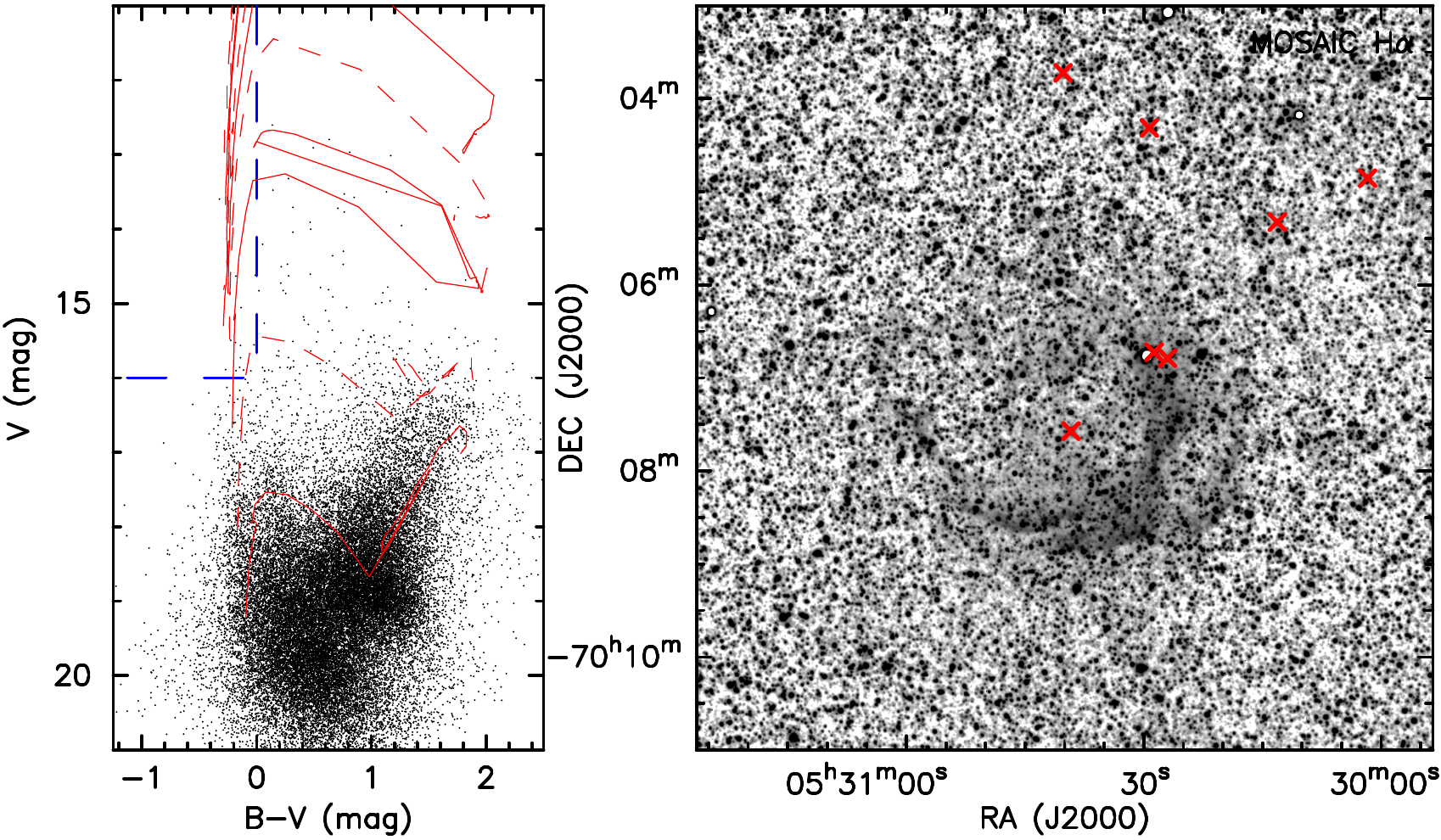}
 \caption{The left panel shows a $B-V$,$V$ colour-magnitude diagram from the MCPS \citep{2004AJ....128.1606Z}. Stellar evolutionary tracks from \citet{2001A&A...366..538L} are shown as dashed red lines (bottom to top: 5, 15, 25 and 60 $M_\odot$) and solid red lines (bottom to top: 3, 9, 20 and 40 $M_\odot$). Heavy dashed blue lines denote the selection criteria for B-star candidates, which lie in the top left corner of the diagram. The right panel shows the MCELS-2 \Halpha\ image of the area around \SNR. B-star candidates are denoted by overlaid red crosses.}
 \label{fig8}
 \end{center}
\end{figure*}

Without reliable polarisation measurements at any radio-continuum frequency we cannot determine the Faraday rotation and thus cannot deduce the magnetic field strength. However, we can use the new equipartion formula derived by \citet{0004-637X-746-1-79} from diffusive shock acceleration (DSA) theory \citep{1978MNRAS.182..443B} to estimate a magnetic field strength: This formula is particularly relevant to magnetic field estimation in SNRs, and yields magnetic field strengths between those given by classical equipartition \citep{1970ranp.book.....P} and revised equipartition \citep{2005AN....326..414B} methods. The average equipartition field over the whole shell of \SNR\ is $\sim$53~$\mu$G (see \citet{0004-637X-746-1-79}; and corresponding "calculator"\footnote{Calculator available at http://poincare.matf.bg.ac.rs/\~{}arbo/eqp/ }), corresponding those of middle-aged SNRs where the interstellar magnetic field is compressed and amplified by the strong shocks.

Fig.~\ref{fig7} shows a surface brightness--diameter ($\Sigma-D$) diagram at 1~GHz with theoretically-derived evolutionary tracks \citep{2004A&A...427..525B} superposed. \SNR\ lies at $(D,\Sigma)$ = (48~pc, $1.1\times 10^{-21}$~W m$^{-2}$~Hz$^{-1}$~Sr$^{-1}$) on the diagram. Its position tentatively suggests that it is in the early Sedov phase of evolution ---  expanding into a very low density environment with the canonical initial energy of a supernova explosion ($10^{51}$ ergs).

High-mass stars rarely form in isolation, so core-collapse supernovae are expected to be associated with other high-mass stars. We used data from the Magellanic Cloud Photometric Survey \citep[MCPS][]{2004AJ....128.1606Z} to construct colour-magnitude diagrams (CMDs) and identify blue stars more massive than $\sim$8~$M_\odot$ within a 100~pc (396\arcsec) radius of \SNR. The CMD in Fig.~\ref{fig8} (left) contains only 13 B-star candidates (V$<$16, B-V$< $0). The red crosses in Fig.~\ref{fig8} (right) shows where the B-star candidates are with respect to \SNR. These criteria would also find stars as late as B2-3 stars. More stringent criteria (V$<$14, B-V$<$0), roughly equivalent to searching for OB stars in the \citet{1970CoTol..89.....S} catalog, would find only 1~star.

Comparison of the star formation histories \citep{2009AJ....138.1243H} in the vicinity of \SNR\ and SNR~J0529--6654 (Bozzetto et al., in press) yields a significant difference: The star formation rate near \SNR\ shows an upturn around 50~Myr ago, whereas the vicinity of SNR~J0529--6654 has a strong spike in the star formation rate in the last 12--25~Myr. The lack of recent high-mass star formation around \SNR\ suggests that it is more likely to be the remnant of a Type~Ia supernova.


\section{Conclusion}
 \label{conclusion}

We have carried out the first detailed multi-frequency study of the LMC \SNR, showing that:
\begin{enumerate}
\item \SNR\ is a relatively large (215\arcsec$\pm$4\arcsec\ $\times$ 180\arcsec$\pm$4\arcsec\ (52$\pm$1 pc $\times$ 44$\pm$1 pc)) shell-type SNR;
\item It has radio spectral index $\alpha=-0.85\pm0.13$ between 843~MHz and 9000~MHz, but the spectrum appears to be peaked/curved;
\item \SNR\ is in the early Sedov phase, expanding into a very low density environment;
\item The average equipartition field over the whole shell of \SNR\ is $\sim$53~$\mu$G;
\item There is a lack of recent local high-mass star formation, suggesting that \SNR\ is the remnant of a Type Ia supernova. 
\end{enumerate} 
With strong optical \SII\ emission (\SII/\Halpha$>$0.4), the presence of non-thermal radio and  X-ray emission, this object satisfies all three criteria for classifying it as an SNR. 


\acknowledgements
We used the {\sc karma} and {\sc miriad} software package developed by the ATNF. The ATCA is part of the Australia Telescope which is funded by the Commonwealth of Australia for operation as a National Facility managed by CSIRO. The Magellanic Clouds Emission Line Survey (MCELS) data are provided by R.C. Smith, P.F. Winkler, and S.D. Points. The MCELS project has been supported in part by NSF grants AST-9540747 and AST-0307613, and through the generous support of the Dean B. McLaughlin Fund at the University of Michigan, a bequest from the family of Dr. Dean B. McLaughlin in memory of his lasting impact on Astronomy. The National Optical Astronomy Observatory is operated by the Association of Universities for Research in Astronomy Inc. (AURA), under a cooperative agreement with the National Science Foundation. This research is supported by the Ministry of Education and Science of the Republic of Serbia through project No. 176005. P. Maggi acknowledges support from the Bundesministerium f\"ur Wirtschaft und Technologie/Deutsches Zentrum f\"ur Luft- und Raumfahrt ( BMWI/DLR) grant FKZ 50 OR 1201.

\bibliographystyle{aa}
\bibliography{aa-0530-7007.F1-R2-Acc}
\end{document}